\documentclass[conference,a4paper]{IEEEtran}
\usepackage{xcolor}
\usepackage{balance}
\usepackage{amsmath}
\usepackage{diagbox}
\usepackage{hhline}

\ifCLASSINFOpdf
  \usepackage[pdftex]{graphicx}
\else
  \usepackage[dvips]{graphicx}
\fi
\ifCLASSOPTIONcompsoc
 \usepackage[caption=false,font=normalsize,labelfont=sf,textfont=sf]{subfig}
\else
 \usepackage[caption=false,font=footnotesize]{subfig}
\fi
\begin{document}
%
\title{3D Printed Discrete Dielectric Lens With Improved Matching Layers}


\author{\IEEEauthorblockN{
Juan Andr\'es V\'asquez-Peralvo\IEEEauthorrefmark{1},   
Jos\'e Manuel Fern\'andez-Gonz\'alez\IEEEauthorrefmark{1},   
Thomas Wong \IEEEauthorrefmark{2},    
}                                        
\IEEEauthorblockA{\IEEEauthorrefmark{1}
Grupo de Radiaci\'on, Depto. de Se\~nales, Sistemas y Radiocomunicaciones, ETSI Telecomunicaci\'on, Universidad Polit\'ecnica\\ de Madrid, 
      Madrid, Spain,   \{jvasquez,jmfdez\}@gr.ssr.upm.es }

\IEEEauthorblockA{\IEEEauthorrefmark{2}
Illinois Institute of Technology, Chicago, IL 60616\\ twong@ece.iit.edu}
 
}



\maketitle

\begin{abstract}
This paper presents a non-zoned discrete dielectric lens comprising two or three matching layers to reduce the 50-110 GHz frequency range reflections. Based on Chebyshev and binomial multi-section transformers, the designed models use matching layers at the top and bottom. In addition, the presented designs use pins instead of the conventional slots for the matching layers, thus easing the manufacturing process. The results show that the broadband realized gain obtained using the proposed design is higher for both the two- and three-layer design than the commonly used quarter-wave transformer. A Binomial lens with two matchings layers using 38 unit cells is fabricated and illuminated by an open-ended waveguide to validate the simulation results obtained using CST Microwave Studio. The fabrication process uses stereolithography additive manufacturing.

\end{abstract}

\vskip0.5\baselineskip

\section{Introduction}
\begin{figure}
\centering 
\includegraphics[width=0.7\columnwidth]{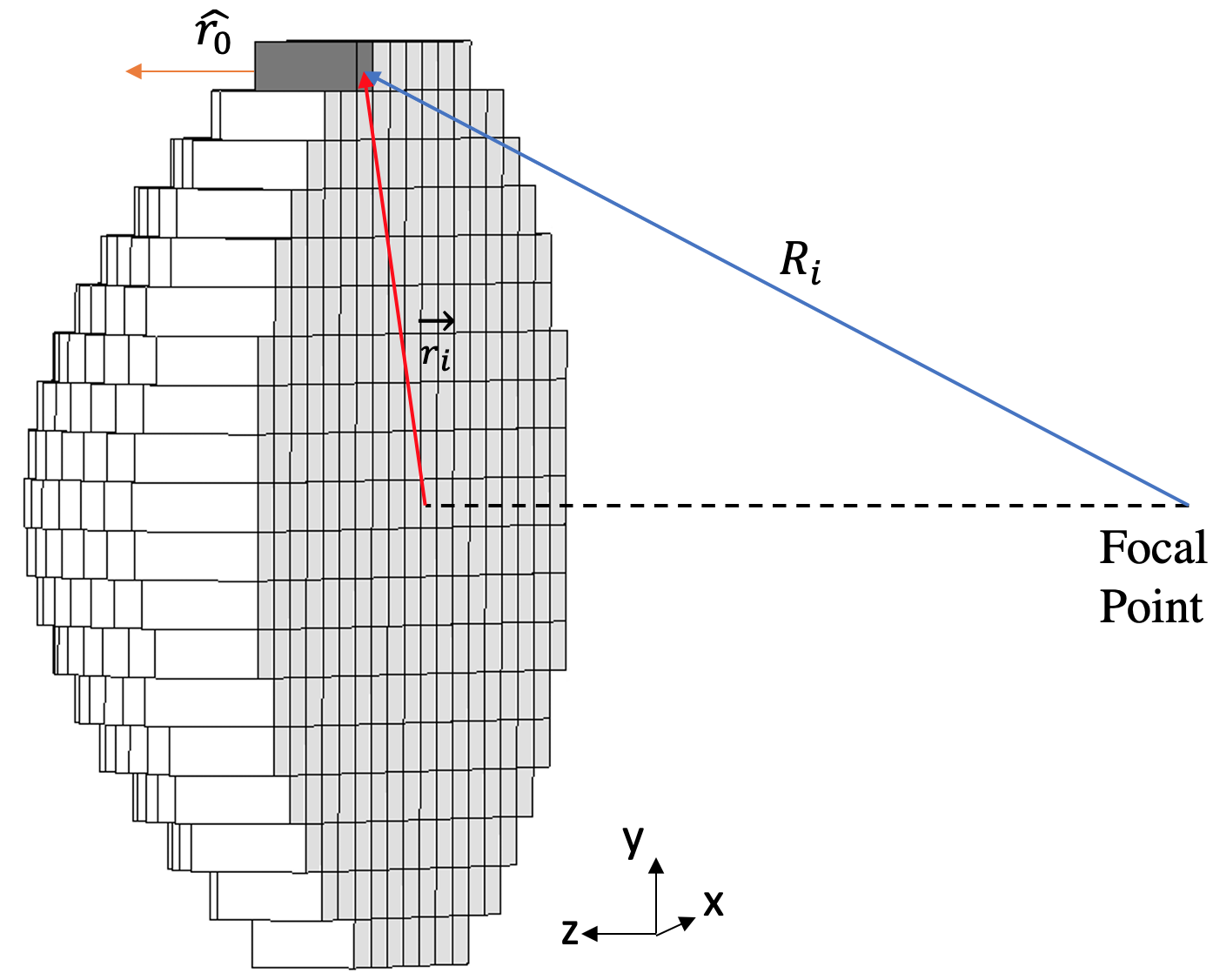}
\caption{Ray tracing of the path length constrain condition, used in a discrete dielectric lens.}
\label{fig:Fig1}
\end{figure}

Lenses have been used to improve the gain of various types of antennas in different applications, including satellite communications, optical measurement tests, automotive communications, to mention a few. The design of lenses is based on the path length constrain condition of Fermat's principle, and based on this, elliptic, hyperbolic, double refractive, zoned, and other lenses can be designed \cite{Lee1988}. A discrete dielectric lens, also known as a dielectric transmit-array, is designed based on the F/D ratio, the source antenna, and the number of unit cells. In turn, the unit cells have other design parameters such as period, the permittivity of the dielectric, and length. The unit cell period has approximately half the wavelength at the center frequency. The permittivity of the dielectric will primarily dictate the thickness of the lens. The higher the permittivity, the smaller the overall thickness of the lens. The length of the unit cell allows the generation of the phase gradients necessary to concentrate the beam to a point or even to beam steering. \cite{7833475}. The phase gradient can be obtained using (\ref{equ:Fermat}), which is graphically illustrated in Fig. \ref{fig:Fig1}.
\begin{equation}\label{equ:Fermat}
\psi_i=k(R_i-\vec{r_i}\cdot \hat{r_0})+\psi_o 
\end{equation}
Where $k$ is the free space propagation constant, $R_i$ is the distance from the focal point to the $i_{th}$ discrete element of the lens, $r_i$ is the position of the $i_{th}$ element from the lens center, and $r_0$ is the main beam direction.\\
Within the literature corresponding to dielectric lenses, different works have been found. In the work \cite{Mateo-Segura2014}, the authors present a planar Luneburg lens using concentric dielectrics of different permittivity. Other authors avoid using different dielectrics by using the same dielectric printed using additive manufacturing, reducing its permittivity by eliminating sections and obtaining spherical \cite{Gbele2014}, extended \cite{Li2016, Biswas2019}, or flat \cite{Monkevich2019, Giddens2020} lenses. The concept of reconfigurability is also proposed by other authors using the concept of flat perforated lenses and an array of \cite{Manafi2019} lenses. None of these works considered the adaptation between the lens medium and air. It is worth mentioning that the concept of dielectric lens broadband is inherent, which is affected when the lens is zoned to reduce its profile. 
There are several works in the literature where discrete zoned lenses are used to increase the gain of various antennas, like a dielectric resonator antenna \cite{Zainud-Deen2015}, a patch antenna \cite{Wang2015}, and a horn antenna \cite{Liu2020}. In the work \cite{Yi2016}, the authors design a discrete zoned lens, working in the V-band, using a quarter-wave transformer to reduce reflections from one medium to another. The quarter-wave transformer is achieved by drilling quarter-wave thick holes into the discretized elements on both faces of the lens. Such that the dielectric constant in that section is equal to $\sqrt{\epsilon_{r_0}\epsilon_{r_m}}$, where $\epsilon_{r_0}$ and $\epsilon_{r_m}$ are the relative dielectric constants of the vacuum and the lens, respectively. One problem with using the quarter-wave transformer is its limited bandwidth, further constrained by the limited frequency range offered by zoned lenses. On the other hand, in \cite{Varela2018}, the authors propose a similar concept as above but for a non-zoned and non-discrete lens. In this case, the antenna gain is reduced by the short bandwidth of the quarter-wave transformer.\\
This work presents an alternative to improve the matching of discrete dielectric lenses in the 50-110 GHz frequency range. This work can also be used for zoned dielectric lenses where reconfigurability can be applied. This design can be used for FSSs measurement setup to reduce reflections between the lenses and radiating antennas. Additionally, the benefits of additive manufacturing for the designed lens prototype are explored by manufacturing a lens using one of the proposed matching layers.
\section{Analysis of multi-section transformers}
Both transformers described below are based on the theory of small reflections between different medium permittivity. In addition, the lengths of each section in both transformers are set to quarter-wavelength. The quarter-wave transformer is not analyzed since it can be obtained by setting the $N$=1 in either matching transformer.
\subsection{Binomial multi-layer matching transformers}
The coupling obtained with this transformer will depend on the number of matching layers added between the two mediums to be coupled. The frequency response accomplished by this coupling is flat, meaning that it has no ripple within the desired bandwidth. The dielectric constants of each section are obtained through each of the sections' characteristic impedance, which depends on the binomial coefficients described in (\ref{equ:Binomial_Coefficients}).
\begin{equation}\label{equ:Binomial_Coefficients}
C_n^N=\frac{N\,!}{(N-n)\,!n\,!}
\end{equation}

Where $N$ is the total number of sections, and $n$ is an integer number, which starts in 0 and ends at $N-1$. The impedances of each medium are calculated using the recursive formula described in (\ref{equ:Binomial_Impedances}).
\begin{equation}\label{equ:Binomial_Impedances}
\ln{\frac{Z_{n+1}}{Z_n}}\simeq2^{-N}C_n^N\ln{\frac{Z_F}{Z_0}}
\end{equation}

Where $Z_0$ is the impedance of the initial medium, $Z_F$ is the impedance of the final medium, $N$ is the total number of sections, $n$ is an integer number, which starts in 0 and ends at $N-1$, and $C_n$ are the section coefficients. 
Finally, the permittivities in each of the media can be calculated by means of (\ref{equ:Binomial_Permitivities}).
\begin{equation}\label{equ:Binomial_Permitivities}
\epsilon_i=\left(\frac{377}{Z_i}\right)^2
\end{equation}

\begin{figure}
\centering 
\includegraphics[width=0.8\columnwidth]{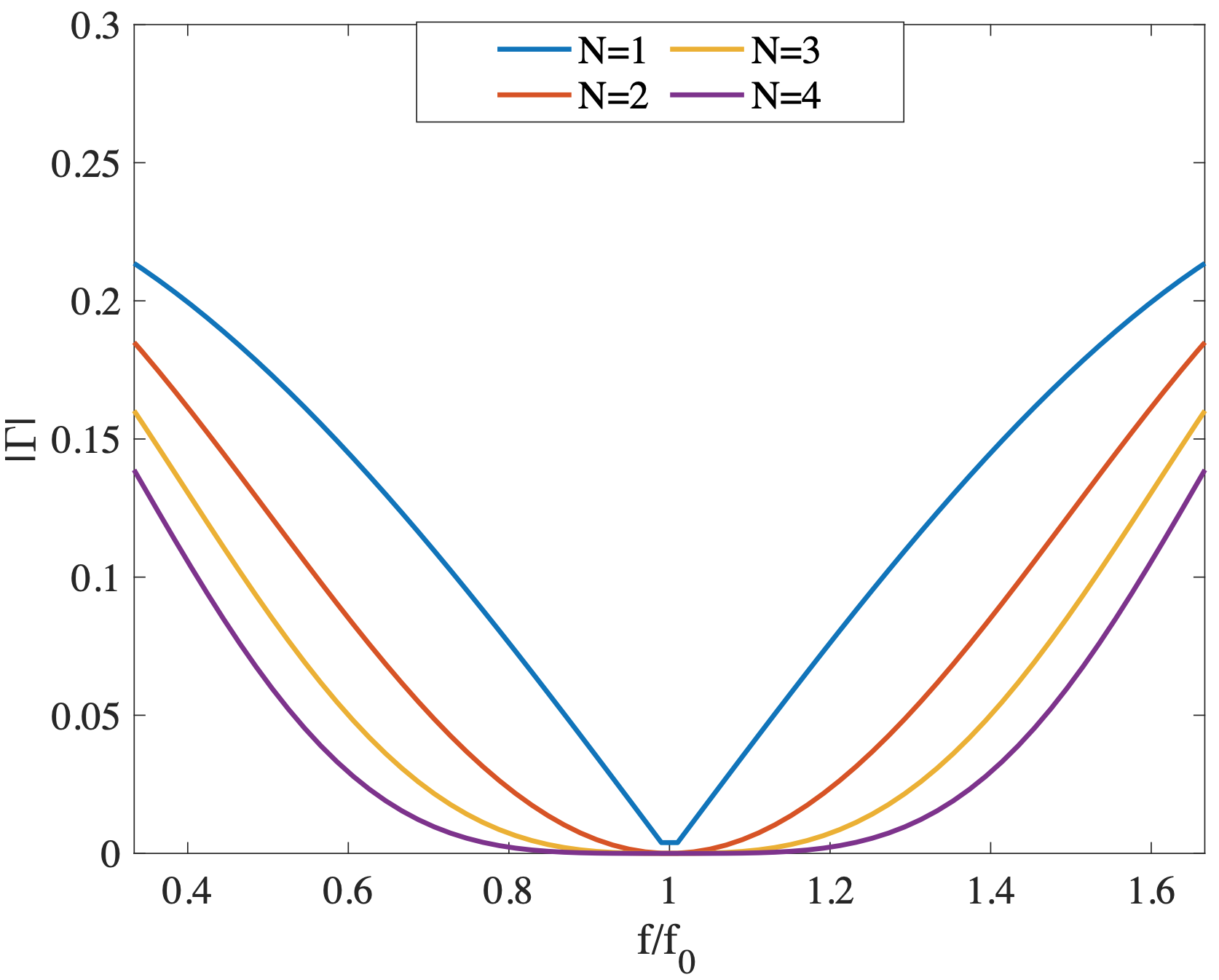}
\caption{Return loss of a binomial multi-section matching transformer. The frequency is normalized to the center frequency and up to four sections are analyzed.}
\label{fig:Fig2}
\end{figure}

The normalized frequency response up to four sections is presented in Fig. \ref{fig:Fig2}. It can be seen that, as the number of sections increases, the response becomes more flat, offering a better passband.
\subsection{Chebyshev multi-section transformer}

\begin{figure}
\centering 
\includegraphics[width=0.8\columnwidth]{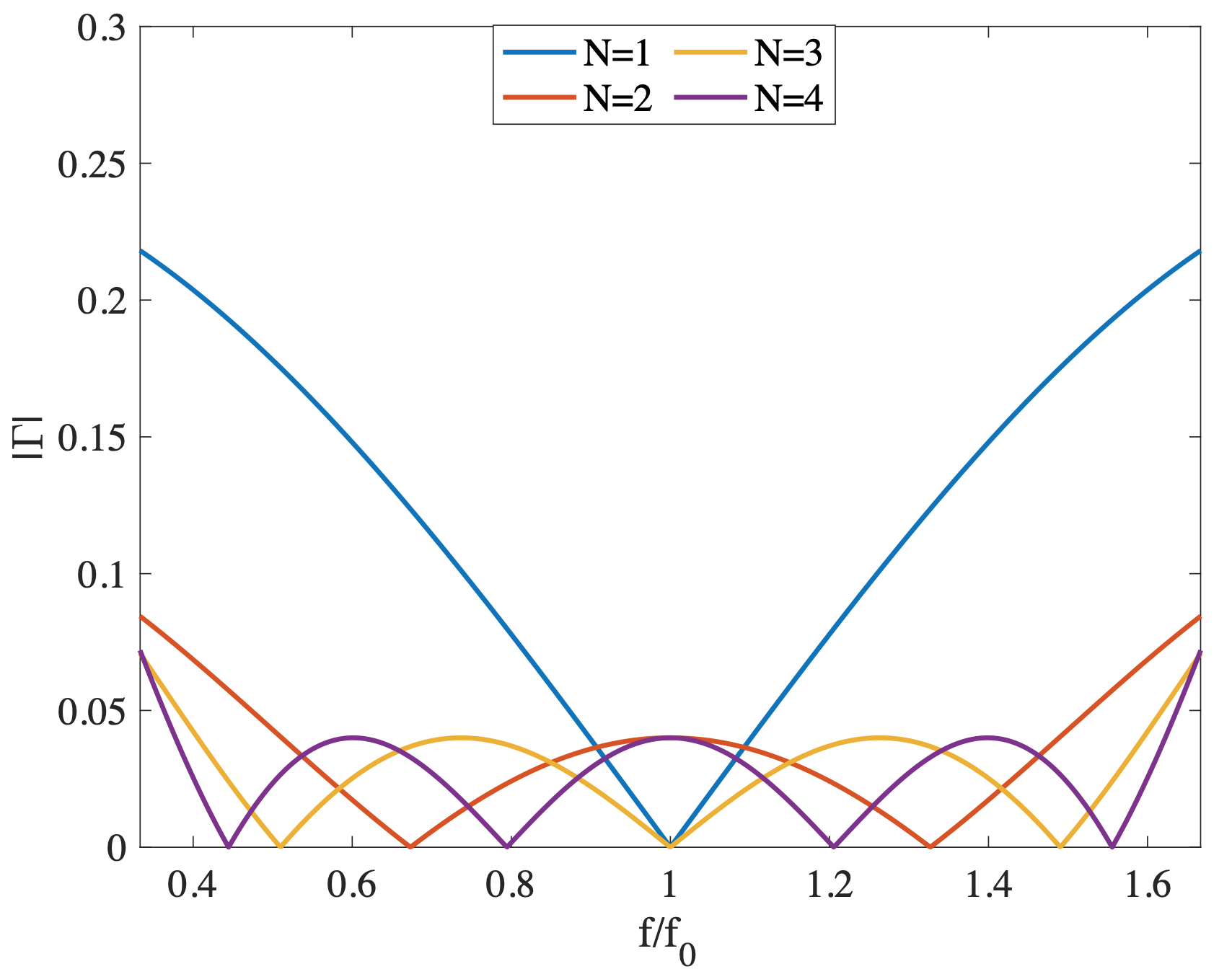}
\caption{Return loss of a Chebyshev multi-section matching transformer. The frequency is normalized to the center frequency and up to four sections are analyzed.}
\label{fig:Fig3}
\end{figure}

In contrast to the binomial matching transformer, the Chebyshev allows choosing the desired bandwidth based on the maximum ripple in the passband $\Gamma_m$. The Chebyshev transformer can be obtained by equating the overall reflection of the matching layers to the Chebyshev polynomial. Since the equations to calculate the matching layers permittivities are different for each $N$ value, we invite the reader to refer to \cite{Pozar2012} for further details. The normalized reflection coefficient of a Chebyshev transformer with up to 4 sections is illustrated in Fig. \ref{fig:Fig3}. As can be seen, the passband depends on the specified ripple and the number of sections.

\section{Multi-section transformer unit cell design}

The unit cell is designed at a center frequency of 80 GHz with a maximum bandwidth $BW_{-20dB}$=60 GHz. The substrate used is the Gray resin which has an $\epsilon_r$=2.738 and a $\tan{\delta}$=0.01 @40 GHz. Based on the previous information, the most suitable period is $p$=1.6 mm.\\
Below are the steps to design the unit cell using quarter-wave, binomial, or Chebyshev transformers.
 
\subsection{Transformer Calculation}
The lengths and permittivities of the impedance transformer sections are designed based on the formulas described in section II. The design parameters are number of layers $N$, first medium permittivity $\epsilon_{r_0}$, final medium permittivity $\epsilon_{r_F}$, and, for the Chebyshev transformer, the ripple $Rm$.  
\subsection{Calculation of the permittivity based on the section thickness}

\begin{figure}
\centering 
\includegraphics[width=1\columnwidth]{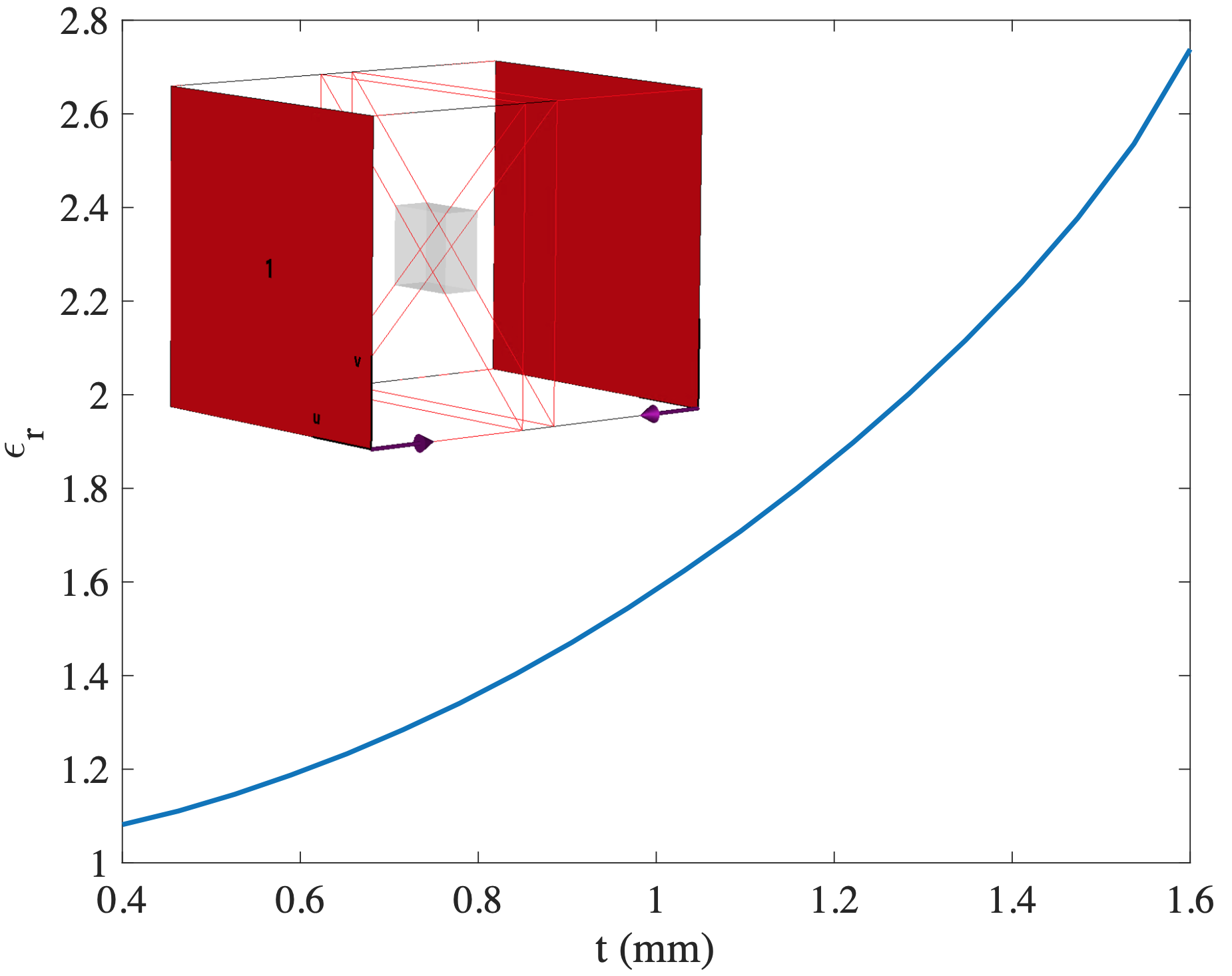}
\caption{Permittivity as a function of the transformer layer thickness. The configuration of the unit cell to obtain the results is also presented.}
\label{fig:Fig3a}
\end{figure}

The permittivities of the transform layers are obtained by reducing the amount of dielectric in a fixed period unit cell. For this purpose, the S-parameters must be obtained for each thickness $t$ variation. The thickness $t$ varies between the values $0<t<p$, while $p$ and the length of the unit cell are kept constant. Then, using a post-processing tool or relating the transmission coefficient phase with the phase constant, the permittivity versus thickness is obtained for a specific frequency. Note that if $t$=$p$ the permittivity is the dielectric permittivity $\epsilon_r$, whereas if $d$$<<$$p$ the permittivity is close to one, as is illustrated in Fig. \ref{fig:Fig3a}.

\subsection{Final unit cell simulation}
\begin{figure}
\centering 
\includegraphics[width=0.8\columnwidth]{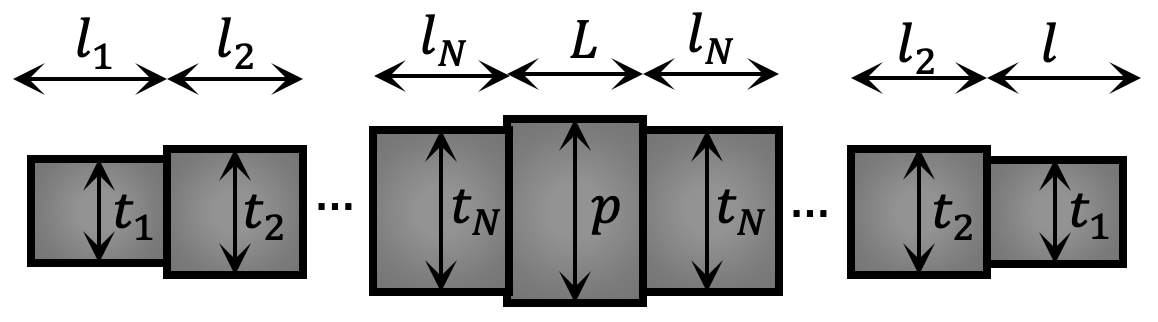}
\caption{Schematic representation of the N-section transformer unit cell. The unit cell has a period $p$ and overall dimensions $2(l_1+l_2+l_3+...+l_N)+L$}
\label{fig:Fig4}
\end{figure}

A total of 6 different unit cells have been designed using the procedure described before. Five use adaptation layers, and one has no adaptation layers. The design parameters and resulting dimensions are shown in Table \ref{Table:Table1}.
\begin{table}
\centering
\caption{Dimensions of the quarter-wave, binomial two- and three-layers, Chebyshev two- and three-layers transformer sections.}
\label{Table:Table1}
\renewcommand{\tabcolsep}{3.5pt}

\begin{tabular}{|c|c|c|c|c|c|c|c|c|} 
\hline
\begin{tabular}[c]{@{}c@{}}\textbf{Matching}\\\textbf{Method}\end{tabular} & \textbf{N } & \textbf{Rm}  & \begin{tabular}[c]{@{}c@{}}\textbf{t1 }\\\textbf{mm}\end{tabular} & \begin{tabular}[c]{@{}c@{}}\textbf{l1 }\\\textbf{mm}\end{tabular} & \begin{tabular}[c]{@{}c@{}}\textbf{t2}\\\textbf{mm}\end{tabular} & \begin{tabular}[c]{@{}c@{}}\textbf{l2}\\\textbf{mm}\end{tabular} & \begin{tabular}[c]{@{}c@{}}\textbf{t3}\\\textbf{mm}\end{tabular} & \begin{tabular}[c]{@{}c@{}}\textbf{l3}\\\textbf{mm }\end{tabular}  \\ 
\hhline{|=========|}
\textbf{Quarter-wave}                                                                       & 1           & \diagbox{}{} & 1.05                                                              & 0.725                                                             & \diagbox{}{}                                                     & \diagbox{}{}                                                     & \diagbox{}{}                                                     & \diagbox{}{}                                                       \\ 
\hline
\textbf{Binomial}                                                                           & 2           & \diagbox{}{} & 0.7                                                               & 0.825                                                             & 1.4                                                              & 0.625                                                            & \diagbox{}{}                                                     & \diagbox{}{}                                                       \\ 
\hline
\textbf{Binomial}                                                                           & 3           & \diagbox{}{} & 0.525                                                             & 0.825                                                             & 1.05                                                             & 0.725                                                            & 1.5                                                              & 0.6                                                                \\ 
\hline
\textbf{Chebyshev}                                                                          & 2           & 0.055        & 0.9                                                               & 0.75                                                              & 1.325                                                            & 0.65                                                             & \diagbox{}{}                                                     & \diagbox{}{}                                                       \\ 
\hline
\textbf{Chebyshev}                                                                          & 3           & 0.01         & 0.575                                                             & 0.85                                                              & 0.875                                                            & 0.725                                                            & 1.475                                                            & 0.6                                                                \\
\hline
\end{tabular}
\end{table}
Additionally, each unit cell will have a section with variable length L, which allows obtaining the phase gradient necessary for the lens design. Each unit cell's initial length with or without matching layers is 6.35 mm. This distance is obtained by adding all the most extended transformer sections (in our case Chebyshev three-layer) and $L$. $L$, as explained before, has a variable length with a minimum value set to 2 mm for mechanical stability reasons. A schematic representation of the unit cell with the matching layers is illustrated in  Fig. \ref{fig:Fig4} 

\begin{figure*} 
\centering 
\includegraphics[width=1.9\columnwidth]{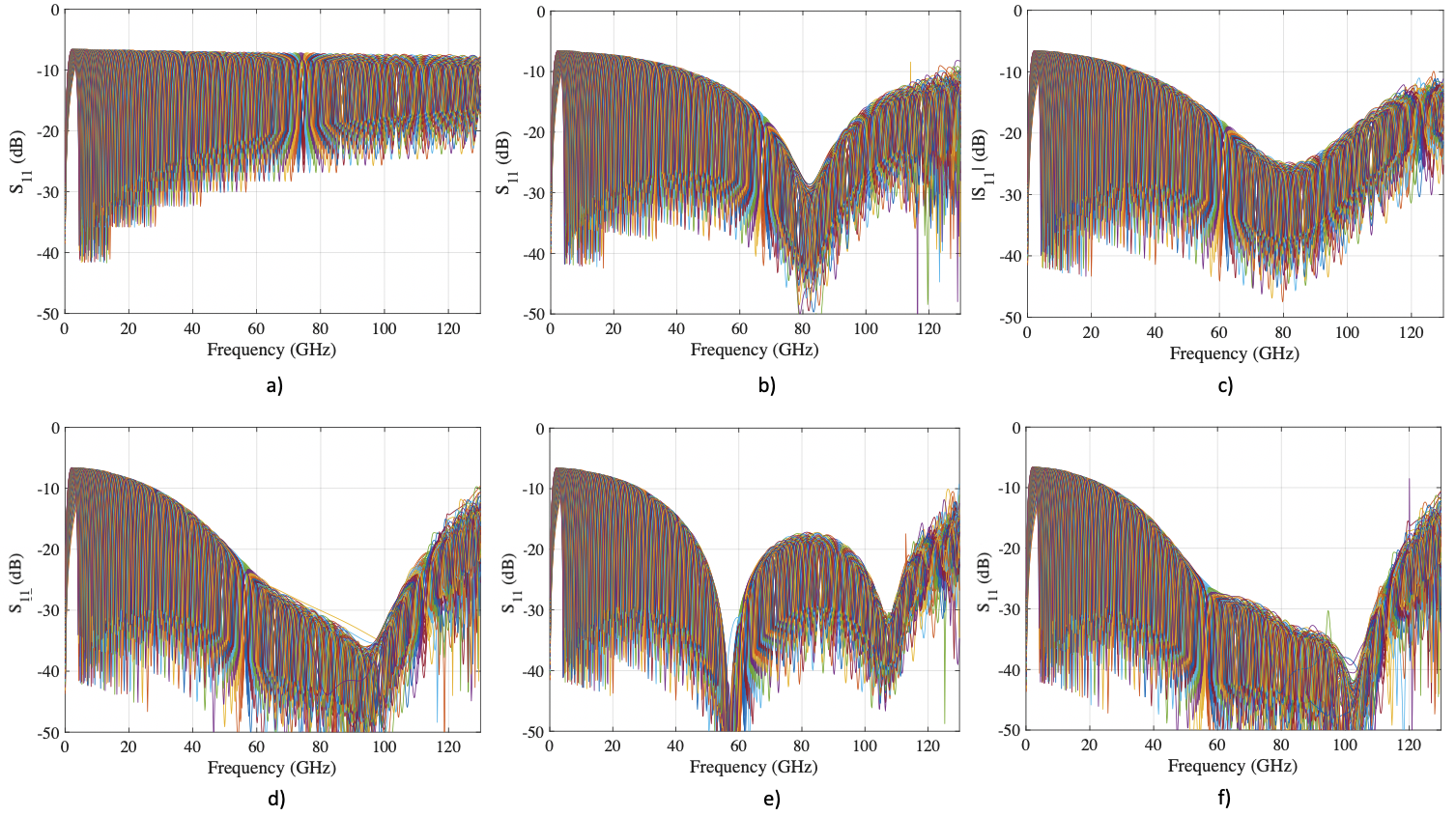}
\caption{Reflection coefficient of different matching transformers. a) Unit Cell with no matching. b) Unit cell quarter-wave transformer c) Unit cell binomial two-sections d) Unit cell binomial three-sections e) Unit cell Chebyshev two-sections f) Unit cell Chebyshev three-sections}
\label{fig:Fig5}
\end{figure*}

The reflection coefficients at different $L$ for the unit cells with and without matching layers are illustrated in Fig. \ref{fig:Fig5}. For the case of the unit cell without matching layers, the reflection coefficient is around -7 dB in the 50 to 110 GHz band. That is, about a quarter of all incident energy is reflected. On the other hand, the quarter-wave transformer presents a 21 GHz passband below -20 dB at a center frequency of 80 GHz. The previous means that using the quarter-wave transformer yields a -20 dB fractional bandwidth of $FBW_{-20dB}$=25\%. In the case of the two- and three-layer Binomial transformer, a fractional bandwidth of $FBW_{-20dB}$=42.5\% and $FBW_{-20dB}$=79.5\% is achieved, respectively. These values go well with the design parameters specified according to the number of matching layers. Finally, the two- and three-layer Chebyshev transformers exhibit fractional bandwidths of $FBW_{-20dB}$=83.45\% and $FBW_{-20dB}$=88.8\%, respectively. Unlike the binomial transformer, the Chebyshev provides a higher bandwidth at the ripple cost in the frequency response, which is very noticeable in the two-layer Chebyshev transformer. It is worth mentioning that the frequency response ripple of the three-layer Chebyshev transformer is not very noticeable due to the small design bandwidth compared to the capabilities of this transformer. 

\section{Lens Design and manufacture}

\begin{figure}
\centering 
\includegraphics[width=1\columnwidth]{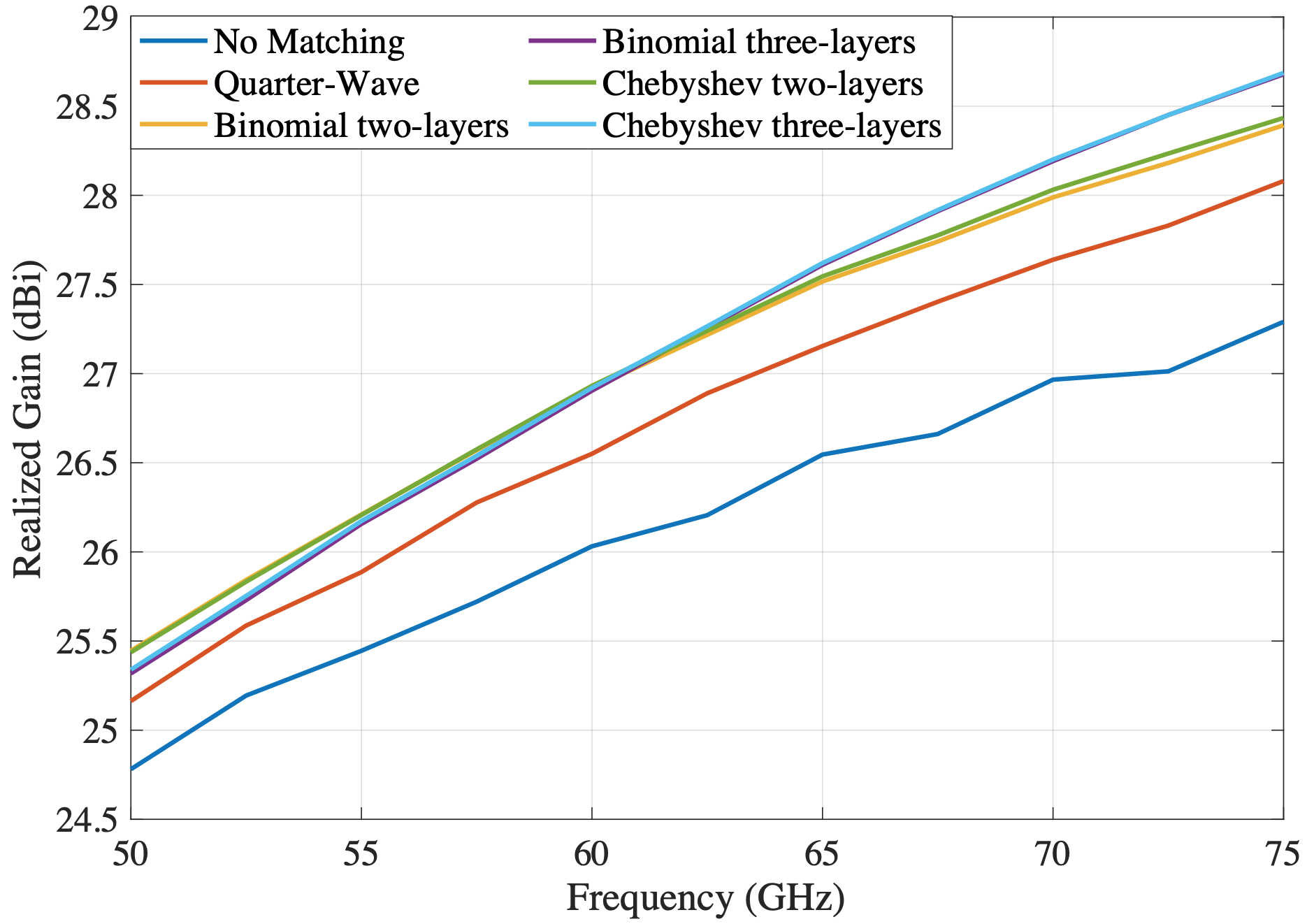}
\caption{Realized Gain simulations corresponding to an OEWG WR-15 with five lenes, each using different matching networks.}
\label{fig:Fig6}
\end{figure}

\begin{figure}
\centering 
\includegraphics[width=1\columnwidth]{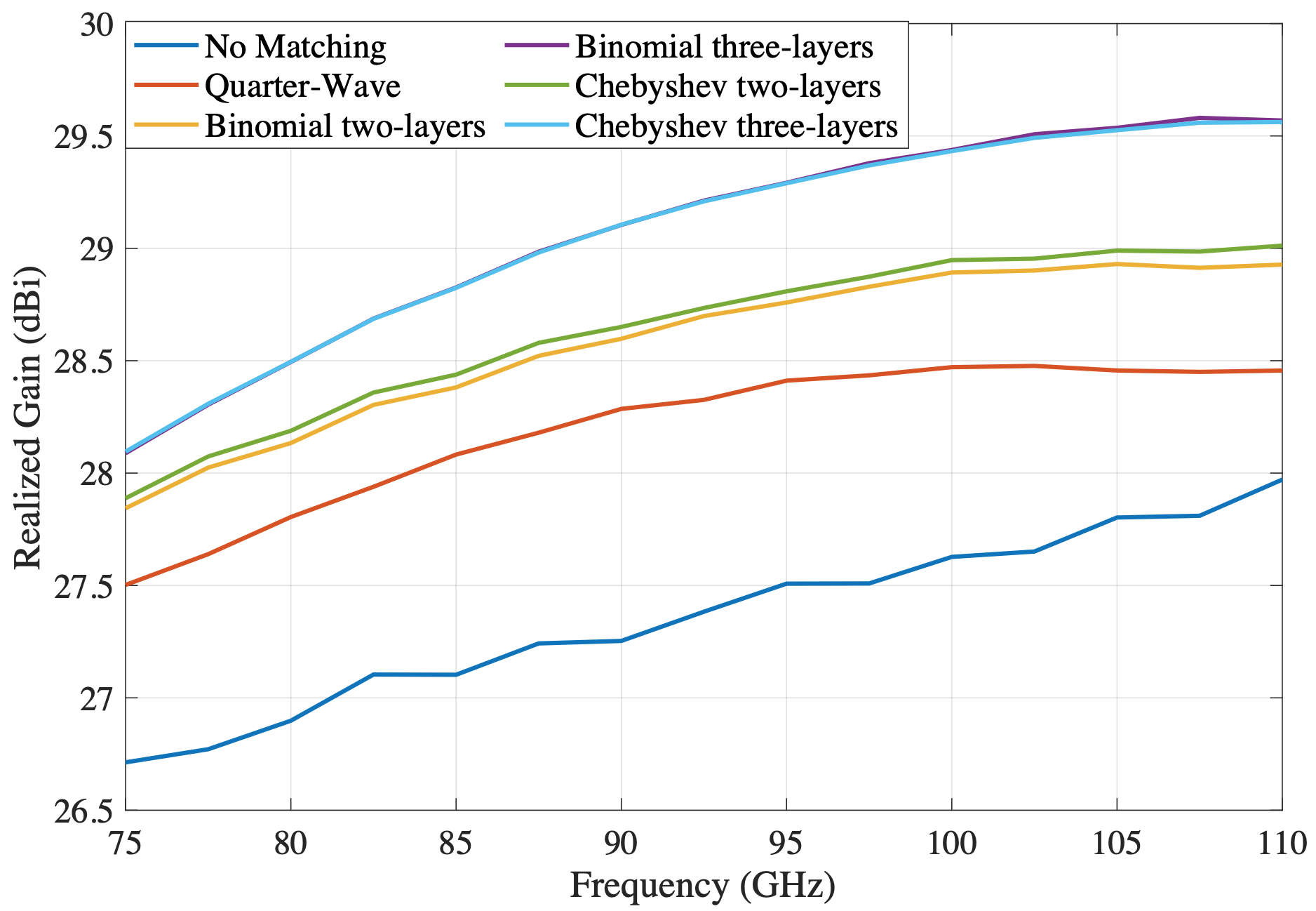}
\caption{Realized Gain simulations corresponding to an OEWG WR-10 with five lenes, each using different matching networks.}
\label{fig:Fig7}
\end{figure}

Six lenses have been designed and simulated using the unit cells described in the previous section. Each lens has $N$=35 and F/D=0.55, allowing for proper illumination and high gain. The antennas to illuminate the lenses are two open-ended waveguides (OEWG), WR-15 and WR-10. Both OEWG will cover the frequency range design of the lens 50 - 110 GHz. Fig. \ref{fig:Fig6} and Fig. \ref{fig:Fig7} show the realized gain results of the six types of lenses for the 50-75 GHz and 75-110 GHz bands, respectively. As can be seen, using the Chebyshev or binomial transformer with two or three sections achieves superior performance to the lens without matching layers or with a quarter-wave transformer. Analyzing the results, using the Chebyshev or binomial transformer with three sections provides excellent performance practically in the entire 50 - 110 GHz band, except for the 50-60 GHz range where its performance is similar to that of the two-section transformers. The previous issue can be overcome by increasing the number of elements in the lens. Another worth mentioning effect is that in the 75-110 GHz band, the realized gain of two- and three-section transformers increases to a maximum of 28.93 dBi @105 GHz and 29.58 dBi @107.5 GHz, respectively. After these frequencies, the gain remains constant up to a specific range and then decays. There are two reasons for this. First, the transformers reach their maximum design passband. The second is that the size of the unit cells at these frequencies would have values greater than $\lambda/2$, which would generate higher modes that would reduce the bandwidth.\\ A binomial lens with two adaptation layers has been manufactured to validate the previously described simulation results. (Results will be given during the presentation of this work).

\begin{figure}
\centering 
\includegraphics[width=1\columnwidth]{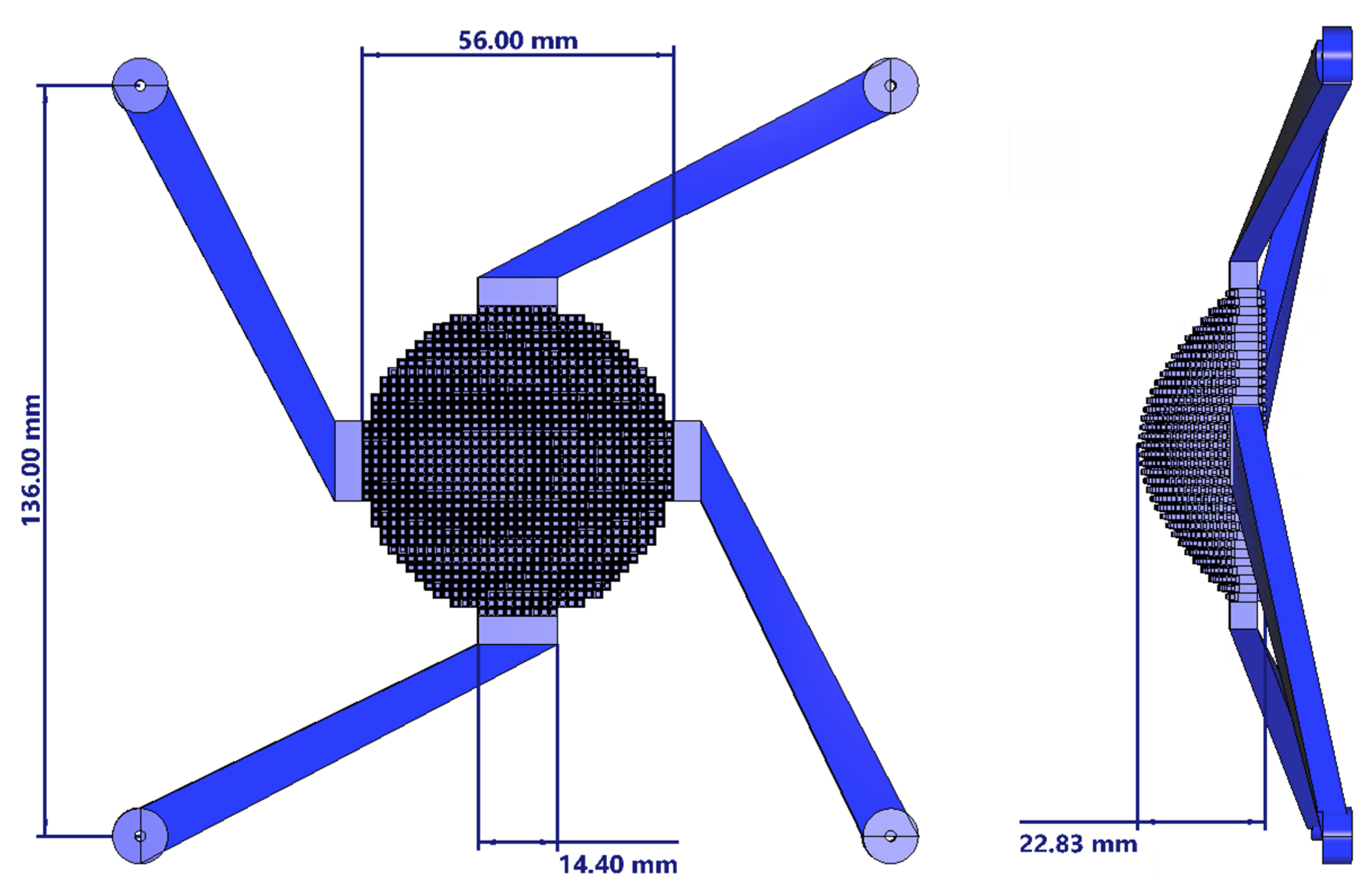}
\caption{Schematic represetation of the proposed lenses using a 2L Chebyshev transfrormer. THe lenses inludes the mechanic support used for measurements. }
\label{fig:Fig8}
\end{figure}

%
%
The prototyping process is performed using additive manufacturing, specifically stereolithography (SLA). Additionally, a bracket has been designed and manufactured for mechanical support, which allows the lens to hold on to the OEWG during measurement, as can be seen in Figure \ref{fig:Fig8}.
\begin{figure}
\centering 
\includegraphics[width=1\columnwidth]{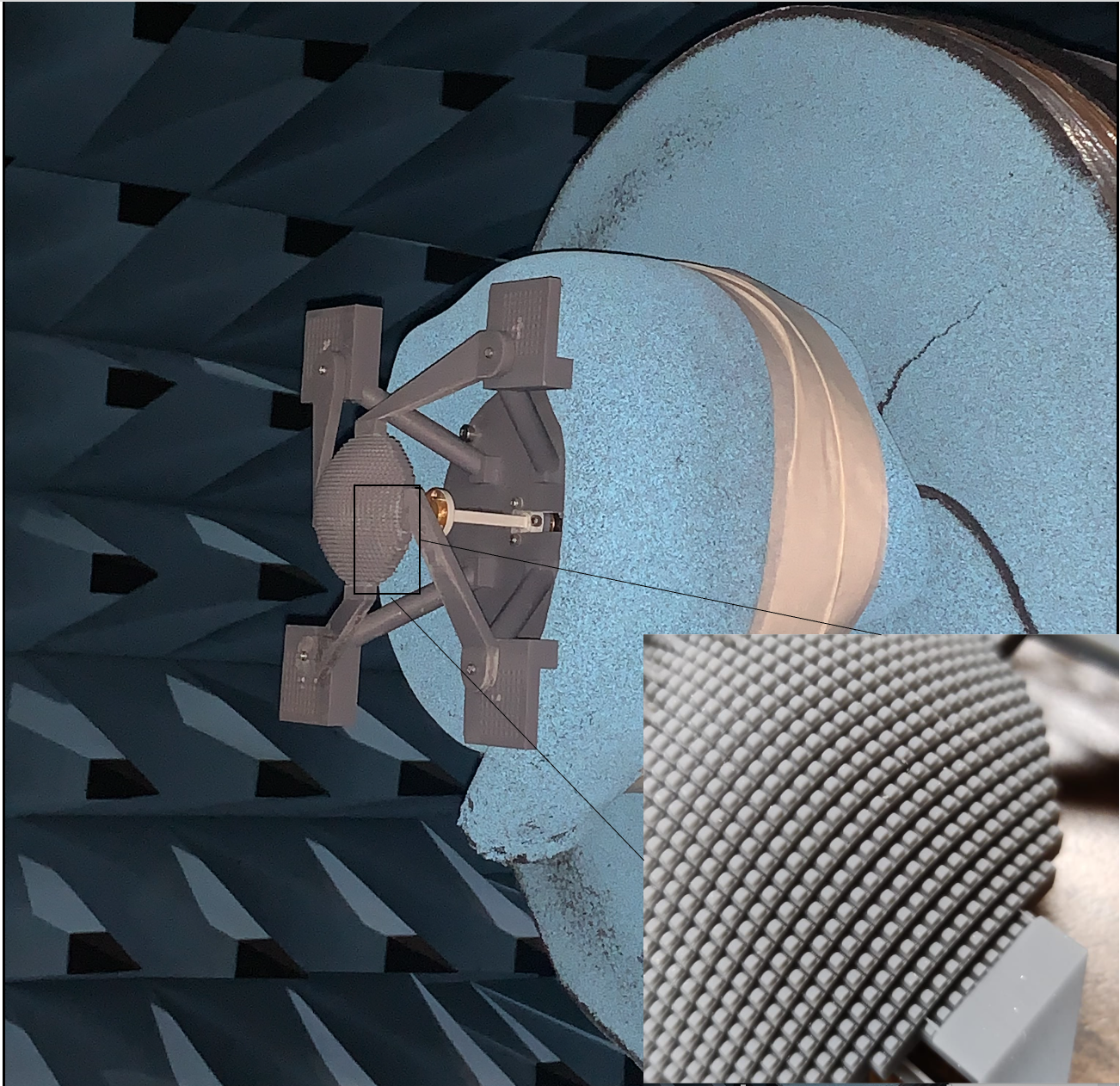}
\caption{Actual photo of the proposed lenses using Chebyshev 2L transformes during measurement. The lens has a total of 36 unit cell.}
\label{fig:Fig9}
\end{figure}

\section{Conclusion}
It has been shown that using multi-section matching transformers can reduce the losses of a discrete lens over a large bandwidth. Depending on the transformer, different passbands can be obtained. For the case of the binomial transformer, the passband is a function of the number of matching layers, while for the Chebyshev transformer depends not only on the matching layers but also on the ripple. Finally, these lenses can be quickly and cheaply manufactured using additive manufacturing techniques.

\section*{Acknowledgment}

This work was supported by the Spanish Government, Ministry of Economy, National Program of Research, Development and Innovation under the project New Array Antenna Technologies and Digital Processing for the FUTURE Integrated Terrestrial and Space-based Millimeter Wave Radio Systems - UPM-InTerSpaCE (PID2020-112545RB-C51).



\bibliographystyle{IEEEtran}
\bibliography{Lens_Antennas.bib}
%
%
%
%
%
%
%
%
%
%
%
%

\end{document}